\documentstyle[amsmath,twoside,fleqn,mont,epsfig,epsf]{article}

\newcommand{\db}{de$\,$Broglie}
\newcommand{\dbb}{de$\,$Broglie-Bohm theory}
\newcommand{\qeh}{quantum equilibrium hypothesis}
\newcommand{\qm}{quantum mechanics}

\begin{document}
\title{Why isn't every physicist a Bohmian?}
\author{Oliver Passon\\
Zentralinstitut f\"ur angewandte Mathematik\\
Forschungszentrum J\"ulich\\
52425 J\"ulich, Germany\\
email: O.Passon@fz-juelich.de }
\begin{abstract}
This note collects, classifies and evaluates common criticism against the \dbb, including Ockham's razor, asymmetry in the \dbb, 
the ``surreal trajectory'' problem, the underdetermination of the \dbb\ and the question of relativistic and quantum field theoretical
generalizations of the \dbb. We argue that none of these objections provide a rigorous disproof, they rather highlight
that even in science theories can not solely be evaluated based on their empirical confirmation. 
\end{abstract}
\maketitle


\section{Introduction}
In a previous article \cite{how_to} we have argued that the \dbb\ can play an important role in teaching 
quantum mechanics since it provides an alternative view-point and illustrates the peculiar features of quantum phenomena.
Of course most adherents of the \dbb\ would assign a more ambitious meaning to the theory  and do rather claim its 
superiority to the ordinary formulation (or interpretation) of quantum mechanics.
In this note we will examine some common objections raised against the \dbb.

In answering the question posed in the title one should first remark that the ongoing debate on the foundations of 
quantum mechanics has produced a vast number of different schools and interpretations. 
Presumably, the majority of physicists has lost track of this complex debate 
about the measurement problem, hidden variables, EPR, Bell etc. In a similar context this was strikingly 
expressed by David Mermin:
\begin{quote}
Contemporary physicists come in two varieties. Type 1 physicists are bothered
by EPR and Bell's Theorem. Type 2 (the majority) are not, but one has to
distinguish two subvarieties. Type 2a physicists explain why they are not
bothered. Their explanations tend either to miss the point entirely
(like Born's to Einstein) or to contain physical assertions that can be shown
to be false.  Type 2b are not bothered and refuse to explain why.
(quoted from \cite{straumann})
\end{quote}
Even if one takes this remark with a pinch of salt, Mermin's observation that many physicists do not have a well founded 
standpoint in  this affair seems to be correct.

Another reason for being rather indifferent to the \dbb\ is evidently the following. This theory can be viewed as 
a way to solve the conceptual problems of quantum mechanics. Those who are satisfied with the answers given by the standard 
interpretation (e.g. David Mermin \cite{love}) or who favor other non-standard interpretations (like many-worlds, 
consistent histories, Floyd's trajectory interpretation or the like) are consequently not attracted by the \dbb.

However, our concern is with criticism and objections which are explicitly directed against the \dbb.
It is rather popular among adherents of the \dbb\ to blame mainly historical and sociological reasons for the contempt 
of their theory \cite{marabeller,cushing_qm,holland}. We do not negate that these reasons may have played some role, 
although such a claim is hard to verify explicitly. In any case such a position renders the criticism as completely 
irrational and makes a sober discussion difficult.
In fact there has been response to e.g. Bohm's paper from 1952. 
Wayne Myrvold, who has analyzed early objections against the \dbb, writes \cite{early} :
\begin{quote} 
Bohm's theory did not meet with the acceptance in the physics community that Bohm had hoped for. It was not, however, 
ignored; several prominent physicists, among which were Einstein, Pauli, and Heisenberg, wrote articles expressing their 
reasons for not accepting Bohm's theory.  
\end{quote}
In what follows we will also explore these early objections\footnote{In fact, not all of the early reactions were hostile.
For example in 1953 Joseph Keller from the 
New York University published a Physics Review paper in which he analyzed the role of probability in Bohm's 
interpretation \cite{keller}. He qualified Bohm's work as an ``interesting interpretation'' of quantum mechanics.}. 

The objections against the \dbb\ can roughly be divided into two classes\footnote{These classes are not completely disjoint and 
serve also the purpose to structure the presentation.}. 
The first applies  meta-theoretical considerations i.e. invokes criteria 
like symmetry or simplicity to discard the \dbb. Section~\ref{meta} is devoted to these arguments.
The other class of criticism seeks for a more textual or theory-immanent debate, like challenging the
consistency or the ability of the \dbb\ to be generalized. This debate will be reviewed in Section~\ref{textual}.

For completeness we will give a brief summary of the \dbb\ in Sec.~\ref{nut}. A thorough discussion of the \dbb\ can be 
found e.g. in \cite{duerr,undivided,cushing_qm,holland,passon}.

\section{The \dbb\label{nut}}

The \dbb\ describes a non-relativistic $N$-particle system by its wavefunction, $\psi$, and the position, $Q_i$, of the  corresponding
quantum objects (e.g. electrons, atoms or the like).
The wavefunction, which is derived from the ordinary Schr\"odinger equation, guides the particle motion 
via the so-called guidance equation:
\begin{eqnarray}  
\label{ge}
\frac{dQ_i}{dt}=\frac{1}{m_i}\nabla_i S(Q_1,\cdots,Q_N)
\end{eqnarray}
Here $m_i$ denotes the mass of particle $i$, $\nabla_i$ is the nabla operator applied to its 
coordinates and $S$ the phase of the wavefunction in the polar representation $\psi=Re^{\frac{i}{\hbar}S}$. 

Since the guidance condition~\ref{ge} is a first-order equation, one initial condition fixes the motion uniquely.
Given a $\rho=|\psi|^2$ distribution as initial positions Equ.~\ref{ge} will reproduce all
predictions of ordinary quantum mechanics in terms of position distributions. Since all measurements can be 
expressed in terms of position (e.g. pointer positions) this amounts to full accordance with all predictions of
ordinary quantum mechanics.  Thereby the \dbb\ assigns a distinguished role to position and does not independently 
assign possessed-values to other observables. This ensures that the Kochen-Specker ``no-go'' theorem does not apply to the \dbb. 
What might be regarded as the values of quantum observables like spin, momentum or the like get established only in the context of 
a corresponding measurement-like experiment. From the viewpoint of the \dbb\ this ``contextuality'' amounts essentially to the 
observation, that the outcome of an experiment depends on the way it is performed.

As mentioned above the \dbb\ reproduces all predictions of ordinary quantum theory 
provided that the initial positions of particles described by the wavefunction $\psi$ are 
$|\psi|^2$ distributed. The motivation of this so-called \qeh\ has been explored for example in \cite{dgz,valentini}. 
Most important, the quantum mechanical continuity equation (Equ.~\ref{ce}) ensures that this condition is 
consistent i.e. any system will stay $|\psi|^2$ distributed if the \qeh\ holds initially. 
\begin{eqnarray}
\label{ce}
\frac{\partial |\psi|^2 }{\partial t}+\nabla  \left ( |\psi|^2 \cdot \frac{\nabla S}{m} \right ) = 0.
\end{eqnarray}
It follows that in a universe being in quantum equilibrium it is not possible to control the initial positions beyond the  
$|\psi|^2$ distribution. Hence the \dbb\  does not allow for an experimental
violation of Heisenberg's uncertainty principle \cite{valentini}.
While ordinary quantum mechanics assumes that probability enters on a fundamental level, the
\dbb\ is deterministic and probability enters only as an expression of ignorance. However, given the \qeh\ this ignorance
holds in principle. Thus the fundamental determinism is turned into predictive indeterminism.

The important feature of Equ.\ref{ge} is its non-locality. The guidance equation links the motion of every particle to the 
configuration of the whole system, no matter how distant its different parts are. Technically expressed this follows from 
the fact, 
that the wavefunction $\psi$ (hence its phase $S$) at a given time is a function on the configuration space ${\rm I\!R}^{3N}$.
It is exactly this non-locality which allows the \dbb\ to violate the Bell inequalities \cite{bell_ungl} 
as demanded by experiment\footnote{Note, that the two ``no-go'' theorems of Bell and Kochen-Specker actually strengthen the 
position of the \dbb. The first shows that ``hidden variable'' theories have to be non-local in order to agree with the 
predictions of \qm. The second rules out that a mapping of all observables to possessed values can be achieved. 
Hence, the corresponding properties of the \dbb\ (namely the non-locality and contextuality) are no shortcomings but 
necessities.}. However, this non-locality vanishes if the wavefunction factorizes in the contributions of the 
different particles. 

The guidance condition~\ref{ge} can be motivated in different ways and its precise status gives rise to different 
interpretations of the \dbb. 
The starting point of Bohm's original presentation of the theory\footnote{We refer to this theory as \dbb\ since 
Louis de$\,$Broglie presented similar ideas already in 1927 \cite{debroglie}. David Bohm's work in 1952 was done independently.} in 
1952 \cite{bohm1} was the 
decomposition of the Schr\"odinger equation for the wavefunction $\psi=Re^{\frac{i}{\hbar}S}$ into a set of two equations 
for the real functions $R$ and $S$. The resulting equation for $S$ has a structure similar to the classical Hamilton-Jacobi 
equation for the action $S$, which implies $p=\nabla S$. The only difference is the appearance of an extra term which Bohm 
named ``quantum potential'':
\begin{eqnarray}  
\label{qp}
U_{\mathrm{quant}}= - \frac{\hbar^2}{2m} \frac{\nabla^2 R}{R}
\end{eqnarray}
Bohm (and later also e.g. Hiley \cite{bh} and Holland \cite{holland}) regard the quantum potential as the key ingredient 
of the \dbb\ and  derive all its novelty from it. The guidance equation~\ref{ge} is only viewed as a  ``special assumption'' 
\cite{bohm2} or a ``consistent subsidiary condition'' \cite{bohm3}.

In contrast to this position another school of the \dbb\ regards the guidance condition as the fundamental equation and avoids
emphasizing the quantum potential. The main proponents of this school are  D\"urr et al. \cite{duerr,nor,dgz,duerr97} 
who have named their version of the \dbb\ ``Bohmian mechanics''. This view was anticipated by John Bell in his work on the \dbb\
\cite{speakable}. 
In fact the guidance equation can 
be motivated without appeal to the Hamilton-Jacobi equation from symmetry arguments alone \cite{dgz}. According to this 
position the quantum potential deserves no special attention and is rather viewed as an artefact which enters the discussion 
when the classical limit of the theory is treated. 

One should not mistake this discussion as only quibbling over a mathematical ambiguity in the formulation of the theory. 
In fact these different interpretations of Equ.\ref{ge} are related to a substantial difference of view on e.g. the role of 
``observables'' 
other than position or the meaning of the wavefunction. Our discussion of objections against the \dbb\ is in part complicated  
by this debate 
on the interpretation. If some criticism applies more strongly or solely to one specific interpretation of the \dbb, it does not 
undermine the concept as a whole. Likewise the different interpretations provide different replies to the objections.
The different interpretation of the \dbb\ will be disentangled elsewhere \cite{int_of_bm}.

Since the rest of our note will be concerned with the objections against the \dbb\ we should balance the discussion
by some brief remarks on its merits.  The supporters of the \dbb\  emphasize its ``clear ontology'' i.e. that the   
vague notion of ``complementarity'' and wave-particle duality becomes dispensable. Within the \dbb\ one can consistently entertain 
the notion of particle trajectories. However, this should not be misunderstood as adherence to classical prejudices but
provides an elegant solution of the measurement problem. The superposition of the wavefunction at the end of a measurement 
causes no difficulty since the actual position of the system corresponds to the actual outcome.
In addition the \dbb\ provides means to deal non-ambiguously  with the question of tunneling time or time-of-arrival 
\cite{gruebl,leavens,leavens2}. Some authors also suggest that the \dbb\ has conceptual advantages over quantum mechanics in connecting 
quantum mechanics to other theories such as chaos theory and classical mechanics \cite{cushing_bowman} or when dealing with CP 
violation \cite{home}.


\section{The meta-theoretical debate \label{meta}}

Most authors accept that the \dbb\ and ordinary non-relativistic quantum mechanics make 
identical predictions i.e. that no experiment can decide which one to prefer\footnote{In fact every now and then such an 
experiment is proposed nevertheless. The attempts to construct circumstances in which the predictions of the \dbb\ and 
quantum mechanics disagree are actually pointless since the \dbb\ reproduces all predictions of ordinary quantum mechanics by definition.
Above all, the Schr\"odinger-equation is part of the \dbb\ and the individual trajectories can not be controlled 
beyond the quantum equilibrium.} . 
Even Wolfgang Pauli admitted in a letter to Bohm from December 1951:
\begin{quote}
I do not see any longer the possibility of any logical contradiction as long
as your results agree completely with those of the usual wave mechanics and as
long as no means is given to measure the values of your hidden parameters
(...). \cite[letter 1313]{briefwechsel})
\end{quote}
But this was only a minimal concession to Bohm. In the absence of any new prediction 
the \dbb\ was accused of being not physics but ``metaphysics'' \cite{pauli52}. Heisenberg 
questioned whether the \dbb\ should be regarded as a new theory at all:
\begin{quote}
From the fundamentally {\em positivistic} (it would perhaps be better to say 
{\em purely physical}) standpoint, we are thus concerned not with
counter-proposals to the Copenhagen interpretation, but with its exact
repetition in a different language.  
(quoted after \cite{early})  
\end{quote} 
The Heisenberg pupil von Weizs\"acker reports on a 
course in the winter term 1953/54 in which they discussed also Bohm's work 
\cite{aufbau}: 
\begin{quote}
Unsere \"Uberzeugung, da\ss\  alle diese Versuche falsch seien, wurde durch das
Seminar best\"arkt. Aber wir konnten uns nicht verhehlen, dass der tiefste Grund
unserer \"Uberzeugung ein quasi \"asthetischer war. Die Quantentheorie
\"ubertraf alle Konkurrenten in der f\"ur eine ``abgeschlossene Theorie''
kennzeichnenden einfachen Sch\"onheit.\footnote{This course strengthened our conviction 
that all this trials were false. But we could not conceal to ourselves that the deeper 
cause for this belief was quasi ``aesthetical''. Quantum mechanics surpassed
all competitors by its simple beauty which characterizes a ``complete theory''.
(translation by the author)}
\end{quote}

However, the above quoted passages alone do not constitute any reason to reject the
\dbb. In the absence of any 
``logical contradiction'' (Pauli) and while objecting to the mere ``repetition in a 
different language''  (Heisenberg) one needs to specify why the ordinary quantum theory 
actually ``surpasses all competitors'' (v. Weizs\"acker). Or to put it differently: 
additional criteria need to be formulated which help to distinguish these theories.

In what follows we collect and evaluate a number of these additional criteria which 
have been suggested by the above mentioned authors and others to underpin their rejection. 
We classify them as ``meta-theoretical'' since they are largely based on requirements which are supposed
to apply to physics theories in general.

\subsection{Ockham's razor}
The obvious objection against the \dbb\ is that it does not make any new predictions while 
postulating the particle-position as a new entity.
If two theories are equivalent the one should be 
preferred which needs fewer premises. Likewise additional premises which do not 
enrich the explanatory power should be removed by invoking ``Ockham's razor''. 
Given this widely accepted principle, it appears natural by some to discard the \dbb\ since the 
particle trajectories seem to be exactly such an extra premise. However, this conclusion can be 
challenged by the following consideration: 
the \dbb\ supplements ordinary quantum mechanics by an equation-of-motion for the 
quantum-particles, but eliminates the postulates 
which are related to the measurement process (not to mention 
how (un-)compelling these postulates are). 
Furthermore the \dbb\ provides a completely new interpretation of quantum phenomena in 
which e.g. probability plays no fundamental role. In other words: the descriptive content
is identical but one can question whether these theories are equivalent at all.
Hence, it is questionable whether the precondition for applying Ockham's razor is met.

\subsection{Asymmetry in the \dbb\label{as}}
Pauli and Heisenberg based their rejection of the \dbb\ mainly on its asymmetry with 
respect to position and momentum \cite{early}. In the absence of any new prediction they 
did not accept this sacrifice. 

The reply to this objection is twofold: (i) The \dbb\ gives position a different ontological status than all other ``observables'' 
\cite{nor} in order to achieve a clear ontology and to solve conceptual problems of the ordinary formulation of \qm. 
After all symmetry is no end in itself. (ii) Moreover, the Hamiltonian in orthodox quantum theory is not invariant under general  
unitary transformations, though it is under the usual space time symmetries. Thus even there it is not 
the case that all observables are on the same footing.



However, in reply to this criticism Hiley and Brown \cite{hiley00,hiley_neu} explore the possibility 
to formulate a Bohm-like theory in other than the position representation. 
Bohm himself took this objection very seriously and was lead to the following modest claim:  
\begin{quote} 
Heisenberg shows that he perhaps did not appreciate that the only purpose 
of this phase of the work was to show that an alternative to the Copenhagen 
interpretation is at least logically possible. 
(D. Bohm, quoted from \cite{early}) 
\end{quote} 

In fact the \dbb\ shows another asymmetry, namely with respect to the wavefunction. 
While the wavefunction acts on the particle position, the particles do not react on 
the $\psi$-field. It is determined independently by the Schr\"odinger equation.
It is true that this constitutes a peculiar feature of the \dbb. 
In reply to this objection D\"urr et al. \cite{duerr97} have suggested that the role of the wavefunction within the \dbb\ 
should be regarded  as analogous to the role of the Hamiltonian in classical mechanics. They state \cite{duerr97}:
\begin{quote}
We propose that the reason, on the universal level, that there is no action
of configurations upon wavefunctions, as there seems to be between all
other elements of physical reality, is that the wavefunction of the
universe is not an element of physical reality. We propose that the wave
function belongs to an altogether different category of existence than that
of substantive physical entities, and that its existence is nomological rather
than material. We propose, in other words, that the wavefunction is a
component of physical law rather than of the reality described by the law.
\end{quote}
In \cite{gold2002} this idea is applied in the context of quantum gravity. 

\subsection{Return to classical physics? \label{return}}
A rather unspecific but never the less common objection against the \dbb\ is its supposed return to classical notions. For example 
Englert states in \cite{rezension} about the purpose of the \dbb\ and its trajectories: 
\begin{quote}
Mit Berufung auf diese Bahnen sind atomare Vorg\"ange dann deterministisch, und das erspart uns die Trauerarbeit, die uns der 
Verlust des deterministischen Newton-Maxwellschen Weltbildes abverlangt.\footnote{With appeal to these trajectories the atomic 
sequence of events gets deterministic and the  mourning-labor about the loss of the Newton-Maxwell world view gets dispensable. 
(translation by 
the author)}
\end{quote}
This claim of  ``backwardness'' is in itself no strong argument against the \dbb. One needs to add (and explain) at least why this 
``return'' is supposed to be  artificial or needless. However, this argument remains weak, since the \dbb\ possesses so many traits 
which are common to quantum mechanics and completely unclassical (e.g. wavefunction on the configuration space, nonlocality etc.pp.) 
that it does a 
disservice to anyone seeking for a ``return to a Newton-Maxwell 
world view''. Agreed, the features of determinism and `objectivity'\footnote{In the sense of `observer independence'} are `classical', 
but in this respect the \dbb\ is as classical as the theory of relativity. 

\subsection{Departure from established principles}
While Section~\ref{return} mentioned the objection of ``backwardness'' the \dbb\ 
meets also with the contrary criticism. Here the bizarre features of the \dbb\ are the subject of 
discomfort\footnote{Some of the objections which have been mentioned in Sec.~\ref{as} do fit into this category as well.}.

According to the \dbb\ the wavefunction produces an actual physical effect on the particle motion. 
In this respect it may be compared to other physical fields like 
electro-magnetic or gravitational fields. This view was for example held by Bell \cite{speakable}[p.128]. 
This introduces peculiar notions into physics indeed. 
First of all the quantum mechanical wavefunction is defined on the configuration space of the system. 
This is in sharp contrast to any other physical field. The non-locality is closely related 
to this feature and will be the subject of Sec.~\ref{n-l}.
As mentioned above (see Sec.~\ref{as}) D\"urr et al. \cite{duerr97} have proposed that the role of the wavefunction 
within the \dbb\ should be rather compared to the role of the Hamiltonian in classical mechanics. 
The Hamiltonian is a function on the phase space, that is of greater dimension and even more abstract than 
configuration space. Following this suggestion certainly weakens this allegation. Interestingly, in this reading of the \dbb\
the role of $\psi$ is similar to the widespread (``orthodox'') view of the wavefunction as a computational tool. 

Viewing the wavefunction as ``nomological'' rather than ``physical real'' helps also to reply to the objection
that the \dbb\ introduces myriads of ``empty waves'' into the picture of physical reality. These are the branches of the wavefunction 
which do not contain the particle on its trajectory hence do not correspond to the actual state of the system.
Although one can argue that due to decoherence effects these empty branches  
do typically not affect the actual system any more\footnote{It seems to be possible to construct circumstances in which empty 
waves do have subtle effects \cite{hardy,vaidman}. This discussion is closely related to the ``surreal trajectory'' debate and will 
be reviewed in Sec.~\ref{essw-d}.}, this feature remains unaesthetic. However, as mentioned above, viewing the wavefunction as 
analog to the Hamiltonian invalidates this allegation.

Finally, no matter whether based on a physical real or nomological wavefunction, does the dynamics of the \dbb\
posses a very unclassical trait. The effect on the particle motion via the wavefunction is independent of its amplitude. This 
can be seen for example when the quantum-potential formulation is 
used. Since $\psi$ appears in the numerator and the denominator of expression \ref{qp}, $\psi$ and $c \cdot \psi$ lead to the 
same effect. Bohm and Hiley have therefore compared the $\psi$-field to radio waves which 
guide an object like a ship on automatic pilot. Here too, the effect of the 
radio waves is independent of their intensity and depends on their form \cite{bh} only. 
Bohm and Hiley have coined the expression ``active information'' for this sort of 
influence and suggest that the quantum potential is a source of this kind of information.
Whether this radio-wave analogy is just a metaphor or leads to any deeper insight  remains to be seen. 
A critical assessment of especially Bohm's metaphors can be found in \cite{guarini}.

Another intriguing property of Bohmian-trajectories gave rise to a specific
objection from Einstein. Since he was one of the famous antagonists of the Copenhagen 
interpretation it is interesting to note that he did not endorse the \dbb\ likewise. 
In a {\em Festschrift} in honor of Max Born in the year 1953, Einstein discussed a system for which 
the \dbb\ predicts a vanishing velocity. Einstein discussed a particle in a 
box as a specific example but the same behavior appears in any system which is described by 
a real wavefunction like e.g. the energy eigenstates of the 
harmonic oscillator. According to Einstein this vanishing velocity  ``contradicts the well-founded 
requirement, that in case of a macrosystem (i.e. for highly excited states) the motion should agree approximately with 
the motion following from classical mechanics'' \cite{early}. 

However, any measurement on the particle  would need a change in the arrangement (e.g. one side of the box would have to be removed). 
The predicted outcome of any such {\em measurement} of e.g. the particle-momentum would be the same as in ordinary quantum mechanics. 
More generally, the Einstein-objection illustrates, that within the \dbb\ the representation of any system is provided by the 
{\em pair} of wavefunction and position, $(\psi,Q_i)$. To focus on properties of one element only can be misleading\footnote{Einstein's
rejection of the \dbb\ is 
clearly not only related to the problem discussed above.  By now famous is his remark in a letter to Born in 1952 about 
the \dbb\ being ``too cheap'' \cite[letter from 12.5.1952]{bornbriefwechsel}. Squires writes in the same context, that 
Einstein ``was not interested in attempts to `cure' the theory; rather he wanted to look elsewhere, to start 
again'' \cite{squires2}. Squires makes an other insightful remark about the \dbb\ and Einstein's probable reason to reject 
it: ``And it is certainly true that we would not have discovered statistical mechanics by adding small corrections to 
thermodynamics, or by adding hidden variables that were in some way `guided' by the free energy, or some other thermodynamic 
quantity'' \cite{squires2}.}.   

\subsection{Underdetermination in the \dbb}

The \dbb\ reproduces the statistical predictions of ordinary quantum mechanics and underpins them with deterministic and continuous 
particle-trajectories. It is tempting to regard the Bohm-trajectories as the actual motion of the quantum-objects.

However, Deotto and Ghirardi \cite{deotto} have shown, that the \dbb\ is underdetermined i.e. the quantum mechanical current can be 
``gauged'' by a divergenceless vector field ${\bf j}^{\prime}={\bf j}+{\bf a}$ with $\nabla {\bf a}=0$. 
The corresponding  guidance condition ${\bf v^{\prime}}={\bf j}^{\prime}/|\psi|^2$  yields the same statistical predictions, while
the individual trajectories differ from the standard \dbb. Hence it is problematic to regard the Bohm-trajectories derived from 
Equ.~\ref{ge} as  representing the ``actual motion'' of the quantum particles.

In order to sustain the ``ontological status'' of the Bohm trajectories one has to formulate additional criteria which 
restrict the possible value of the vector field ${\bf a}$. E.g. D\"urr et al. \cite{dgz}  motivate the guidance equation \ref{ge} from 
symmetry and simplicity constraints. The argument in \cite{hollandundphi} is based on the assumption that the corresponding problem 
for relativistic 
spin $\frac{1}{2}$ particles has been solved uniquely \cite{holland99}. Holland and Philippidis then derive the guidance equation for 
the non-relativistic limit and result in a equation which contains an additional spin-dependent term. Hence, for spin 0 particles the 
original form is recovered. 

However, this ambiguity of the \dbb\ does not undermine its conceptual value. If the above motivations for the specific guidance 
condition ( Equ.~\ref{ge}) are felt unconvincing, the \dbb\ still provides a proof of principle that the deterministic interpretation 
of quantum mechanics is possible. Since the measurement of individual trajectories is beyond the principle reach of experiments 
one should not put too much emphasize on their particular form anyway.

\subsection{The status of the \qeh}
As mentioned in Sec.~\ref{nut}, the \dbb\ reproduces all predictions of ordinary quantum theory 
provided that the initial positions of particles described by the wavefunction $\psi$ are 
$|\psi|^2$ distributed. One may include this assumption in the very 
definition of the \dbb. Equation \ref{ce} ensures that this postulate is consistent 
i.e. any system will stay $|\psi|^2$ distributed when the \qeh\ holds initially. 

However, introducing the \qeh\ as a postulate provokes the objection that thereby the 
wavefunction gets two distinct and logically independent meanings: (i) as the guiding 
field and (ii) as a probability distribution for the particle position. This double role for the wavefunction looks suspicious and 
unaesthetic. Further more it would remain obscure how random behavior enters into the deterministic \dbb. Finally the 
very meaning of such a postulate would be not clear at all. 

It was therefore among the early efforts of Bohm 
to clarify the status of the \qeh\ and to possibly derive rather than postulate it. 
The paper \cite{bohm2} from 1953 was devoted to this question but could derive the \qeh\ 
only for a limited class of systems \cite{cushing_qm,val-sim}. This problem gave rise to 
the development of a modified version of the theory in 1954 including the effect of 
a stochastic disturbance \cite{bv}. A dynamical explanation of the \qeh\ within the original version of the \dbb\ was also
attempted by Valentini  \cite{valentini}.  

A different approach was developed by D\"urr et al. \cite{dgz}. Their analysis is an elaboration of work of John Bell
and is ultimately rooted in the approach of Ludwig Boltzmann to statistical mechanics.
The starting point is that regarding the \dbb\ as a fundamental theory implies that the behavior of subsystems is determined by the 
``wavefunction of the universe'', $\Psi(q)$, and the corresponding 
configuration.
One is therefore not free to simply postulate that subsystems have wavefunctions and are governed by the \dbb. 
However, applying the \qeh\  to $\Psi(q)$ seems to be physically meaningless since we do not have a sample of universes. 
Thus the following two questions need to be addressed: (i) how to assign a wavefunction to a subsystem and (ii) what is the meaning 
of the \qeh\ when applied to $\Psi(q)$. Finally one can ask how to relate these points i.e. how to
justify the \qeh\ for empirical distributions\footnote{By them we mean relative frequencies obtained from repeated experiments 
on subsystems.}.  

Question (i) leads D\"urr et al. to the introduction of the {\em effective} wavefunction.   
Let $q=(x,y)$ be a decomposition of the configuration of the universe into the variable $x$ of a subsystem and $y$ 
for the rest. D\"urr et al. define the {\em effective} wavefunction, $\psi$, of the subsystem as part of the following decomposition:
\begin{eqnarray}
\Psi(x,y)=\psi(x)\Phi(y)+\Psi^{\perp}(x,y)
\end{eqnarray}
The wavefunction $\psi(x)$ represents the subsystem provided that
the $y$-support of $\Phi(y)$ and $\Psi^{\perp}(x,y)$ is macroscopically distinct and that the actual value of $y$ lies in the 
support of $\Phi(y)$. A typical situation of this kind occurs during a measurement on the system described by $x$ with a 
measuring device that has, at the end of the measurement, a definite value in the support of $\Phi(y)$.

Regarding (ii)  D\"urr et al. argue, that the meaning of the quantum equilibrium distribution $|\Psi(q)|^2$ on the universal level 
is {\em not} probabilistic since we do not have a sample of universes. Instead, it provides a so-called  measure of 
{\em typicality}. The notion of  {\em typicality} (though not the word) was introduced by Boltzmann in justifying the 
second law of thermodynamics. This statement holds because an ``overwhelming majority'' of initial conditions leads to a behavior in 
accordance with the second law \cite{lebowitz}. However, the meaning of ``overwhelming majority'' i.e. a measure on the corresponding set, needs 
to be specified. One important requirement for this measure is that it should be ``equivariant'' i.e. the notion of typicality 
should be independent of time. And in fact, the continuity equation \ref{ce} ensures that the measure  $|\Psi(q)|^2$ is equivariant. 

Finally, and that is the central result of \cite{dgz}, D\"urr et al. can prove that within a ``typical'' universe the \qeh\ holds 
for all subsystems. Hence the typical Bohmian universe -- although deterministic -- gives the {\em appearance} of randomness in 
agreement with \qm.

This justification of the \qeh\ has been questioned e.g. by Dickson \cite{dickson}. He notes that D\"urr at al. have not 
shown that  $|\Psi(q)|^2$ provides the {\em only} equivariant measure. Further more Dickson questions that 
equivariance is a preferred property of measures over the initial distributions at all.  He states \cite{dickson}[p. 123]:
\begin{quote}
Equivariance is a {\em dynamical} property of a measure, whereas the question `Which initial distribution is the correct one?' 
involves no dynamics, nor it is clear why dynamical properties of a measure are relevant. 
\end{quote}
This objection challenges the claim that the \qeh\ can be derived rather than postulated. However, it should be noted that for 
the justification of classical thermodynamic the question of how to derive apparent randomness from deterministic laws is
just as controversial.

\section{The theory immanent debate \label{textual}}
Until now we were mainly concerned with meta-theoretical objections which might be viewed as partially subjective. 
Consequently, some of the above mentioned feature of the \dbb\ have been either used to reject this theory or to praise 
its radical novelty.

An other strategy to disclaim the \dbb\ has been to seek for a more textual debate, e.g. 
challenging its consistency or its ability to be generalized. One might say that these arguments try to refute the \dbb\ 
from ``inside'', hence we have classified them as ``theory-immanent''. Most important is the question whether a 
trajectory-interpretation is sustainable in the relativistic domain. 

A clear-cut  disproof of the \dbb\ would be an experiment in which the 
predictions of the \dbb\ and ordinary quantum mechanics differ while the latter 
is confirmed. In fact every now and then such an experiment is proposed. 
The attempts to construct circumstances in which the predictions of the \dbb\ and 
quantum mechanics disagree are actually pointless since the \dbb\ reproduces  all predictions of ordinary quantum mechanics
by definition. Above all, the 
Schr\"odinger-equation is part of the \dbb\ and the individual trajectories can not be controlled 
beyond the quantum equilibrium. This attempts will not be 
considered further and the interested reader may consult 
\cite{pointless,pointless2,answer}. 

\subsection{The ``surreal trajectory'' objection \label{essw-d}}
In 1992 Englert, Scully, S\"ussmann and Walther (ESSW) challenged the \dbb. They claimed that  Bohm trajectories are not realistic, 
but ``surrealistic''.

The corresponding authors analyze the famous delayed-choice double-slit experiment
invented by Wheeler \cite{wheeler} and discussed in the context of the \dbb\ by Bell \cite{speakable}. 
Before we turn to the actual ESSW argument we will first discuss the original set-up.

\begin{figure}[t]
\begin{center}
\centerline{\epsfxsize=2.5in\epsfbox{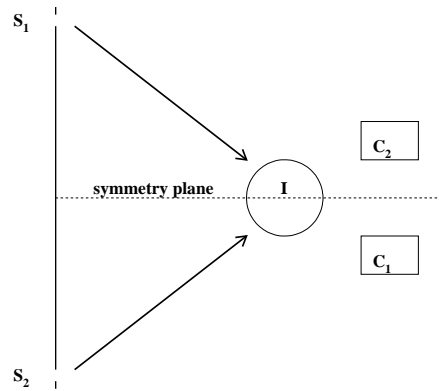}}
\vspace{-1.cm}
\caption[*]{\label{ds} {\em Delayed-choice double slit experiment. One may either detect particles in the detectors $C_i$
or observe interference in region I  by insertion of a screen.}}
\end{center}
\vspace{-0.8cm}
\end{figure}

The delayed-choice double-slit experiment (see Fig.~\ref{ds}) consists of a double slit arrangement in which 
one can freely choose to detect either interference patterns in the region I or particles in the detectors $C_1$ or 
$C_2$ \footnote{i.e. one can insert a screen in region I in order to detect the interference pattern. 
This  choice can be made after the ``particles'' have passed the slits already, hence the name {\em delayed-choice}.}.
The whole arrangement is set up in such a way that by symmetry arguments the trajectories of the \dbb\ are not allowed to cross 
the midplane behind the two-slit screen. They show the ``unclassical'' behavior, that the Bohm-trajectories of the particles hitting the 
upper part behind the screen have traversed the upper slit and vice versa. 

One may modify the arrangement by supplying it
with additional detectors directly behind the two-slit screen in order to investigate 
which slit has been traversed. In such a modified version the interference pattern 
would not occur. Additionally the \db-Bohm trajectories would be allowed to cross the midplane
since, given the degrees of freedom related to the detector, it is no symmetry plane anymore
(see \cite{speakable}[p.111] for the details of this argument.).

The above mentioned extra-detectors directly behind the screen were
assumed to be ``ordinary detectors'', i.e. devices which show a
macroscopic change of state (e.g. pointer positions). We now turn to
the actual ESSW argument.  According to these authors a problem for
the \dbb\ emerges when these extra detectors are chosen to be advanced
quantum optical devices, so-called ``which-way detectors''. These
respond on the transition of single particles without affecting the
translational part of the wavefunction\footnote{E.g. within a
micromaser excited Rydberg atoms can radiate off one photon without
any other significant change of state.}. Again, in the presence of
these devices, we expect the interference pattern to be destroyed. The
special feature of these ``one-bit detectors'' is that even their
excitation does not alter the symmetry (i.e.
$\psi_1(x,y,z)=\psi_2(x,y,-z)$, with $z=0$ being the midplane). Hence
the \db-Bohm trajectories are still forced to ``bounce off'' the
midplane. However, the probabilities $|\psi_1|^2$ or $|\psi_2|^2$ are
in general not confined to one half of the screen.
According to ESSW one arrives at the paradoxical situation that 
the upper which-way detector fires while  the screen is hit below the midplane. 
ESSW conclude:
\begin{quote}
The Bohm trajectory is here macroscopically at variance with the actual, that
is: observed track. Tersely: Bohm trajectories are not realistic, they are
surrealistic.
\end{quote} 
This paper has created a lively debate on the ``surreal trajectory problem'' \cite{barrett,dhs,dfgz,hcm,motc,scully}
and we do not aim at a complete review. 
One objection against the conclusion of ESSW has been their use of the term 
``actual track'' in connection with quantum mechanics.
ESSW try to defend the orthodox interpretation -- but the notion of a ``particle path'' 
is denied within this interpretation. What is meant by ``actual track'' 
is not obvious here. However, ESSW claim that even the observed tracks in a 
bubble-chamber are at variance with the Bohm-trajectories. This would be a serious 
objection against the \dbb\ indeed.

The essential flaw in the reasoning of ESSW is that they consider devices
which are not linked to any macroscopic change of state. This feature is 
crucial because it ensures that their symmetry argument applies.
But given that within the \dbb\ only a change in position (or of the wavefunction)
constitutes a physical fact, such a which-way detector is not regarded as a reliable 
detector for the actual position of the particle on its Bohmian path. 
The additional claim of ESSW, that even the tracks in a 
bubble-chamber differ from  the predicted Bohm-trajectories, is therefore unfounded, since a bubble-chamber does 
convert the excitation into a macroscopic displacement.  

But the situation which has 
been considered by ESSW is a bit more subtle: The authors assume that a macroscopic 
read-out could be connected {\em after} the particle has been finally detected. 
However, it remains true that within the \dbb\ the  which-way device is not regarded as a detector. 
A delayed read-out can not turn it into a more trustworthy device.  The arrangement which has been considered by Englert et al. can 
be viewed as a special case in which
``empty waves'' \cite{hardy} show an effect if they are still coherent. In fact, the non-locality of the \dbb\ makes it 
possible to explain how the which-way detector can be excited even without any trajectory passing through it \cite{dhs,hcm}.
A detailed discussion of how to resolve the ``surreal trajectory problem'' 
within the \dbb\ can also be found in \cite{barrett}. 

Along similar lines also other arguments have been advanced in order to show that ``the Bohmian position does not help to understand the 
result of a measurement'' \cite{ahavaid}. Especially Aharonov et al. \cite{ahavaid,aes,aes2} have 
explored Bohm trajectories in the case of ``weak'' and ``protective'' measurements\footnote{A ``weak measurement'' \cite{wm} is designed 
to change the corresponding system only minimally. A ``protective measurement'' is both, weak and adiabatic \cite{aes}.} 
in order to challenge any ``realistic interpretations of 
Bohm trajectories'' \cite{aes}. Similar to the original ESSW argument these authors construct circumstances in which non-local effects
are exerted i.e.  alleged measuring devices are triggered while the Bohm trajectories do not pass through them. They conclude that 
their analysis 
\begin{quote}
(...) implies that the Bohm trajectories are forever hidden. If you cannot rely on local interactions to determine the `actual position' 
of the particle, then you cannot determine it at all. The concept of position itself becomes shaky. \cite{aes}
\end{quote}
However, Aharonov et al. do not claim the inconsistency of the \dbb:
\begin{quote}
The examples considered in this work do not show that the Bohm's causal interpretation is inconsistent. It shows that Bohmian 
trajectories behave not as we would expect from a classical type model. \cite{ahavaid} 
\end{quote}
Furthermore Aharonov and Vaidman admit, that ``these difficulties follow from our particular approach to the Bohm theory in which the 
wave is not considered to be a `reality'.''

Recapitulating, we note that these investigations have given fascinating insight into detailed aspects of quantum mechanics in general 
and the \dbb\ in particular. They clearly demonstrate that (especially given the exotic measuring devices considered above) the 
trajectories behave completely unclassical and that the \dbb\ is as unintuitive as the usual quantum theory. However, most adherents 
of the \dbb\ never argued that point.

\subsection{Fractal wavefunctions}
A recent argument against trajectory-based interpretations of quantum mechanics in general and the \dbb\ in particular was 
advanced by Hall \cite{hall}. He considers so-called fractal  wavefunctions for which the expression ${\cal{H}}\psi$ is 
divergent\footnote{There are examples for which the expectation value $\langle {\cal{H}} \rangle$ is finite nevertheless 
\cite{hall}.} while the equation  $[ {\cal{H}}-i\hbar\partial_t]\psi=0$ is satisfied still\footnote{The corresponding states are said 
to be solutions of the Schr\"odinger equation in the ``weak'' sense \cite{weak}.}. 
Given that the usual  Schr\"odinger equation  does not hold for these states, Hall argues that the modified Hamilton-Jacobi equation 
can not be derived. Further more, $\nabla \psi$ is not defined and the guidance equation of the \dbb\ fails to provide trajectories for 
these states. 

Given that the corresponding states and their unitary evolution are well defined Hall claims that trajectory-based interpretations 
are at least {\em formally} incomplete. Provided that these states could be actually prepared they may even demonstrate the 
{\em physical} incompleteness.  

In reply to this criticism one may note that the wavefunctions considered by Hall are unphysical. More relevant in this context is the  
question of global existence of Bohmian trajectories. This issue was settled in \cite{eunde}. Only recently Tumulka and Teufel  
\cite{simple} have simplified and extended this proof to the Bohm-Dirac theory. If a wavefunction satisfies the 
conditions for global existence and uniqueness, then it is ensured that it can not evolve into e.g. a fractal state \cite{pc}. 
Hence, Hall's claim about a possible ``physical incompleteness'' seems to be unfounded and his claim of  ``formal incompleteness'' 
amount to no more  than a specific definition of ``formal''. 

\subsection{Non-locality and relativistic generalization \label{n-l}}
The by far most common objection against the \dbb\  is based on its non-locality and its apparent conflict 
with relativity. We will try to disentangle these questions in turn. 

The \dbb\ is explicitly non-local, i.e. the motion of each particle is in general a 
function of the coordinates of the whole system, no matter whether they are space-like separated. This non-locality vanishes only 
if the wavefunction factorizes in the contributions of the different 
quantum objects.
Whether this is viewed as an unacceptable feature depends on the attitude towards 
the problem of non-locality in quantum mechanics in general. In ordinary quantum mechanics the problem of non-locality appears 
in at least two places: (i) violation of Bell-inequalities and (ii) reduction of the wavefunction.

Following the work of Bell \cite{bell_ungl} and the 
experimental confirmation of quantum mechanics in EPR-Bell experiments \cite{aspect} it 
became widely (but not universally \cite{tellus,muynck}) accepted that quantum mechanics itself is 
``non-local''. Following this opinion the non-locality allegation against the \dbb\ seems to be even completely
groundless. However, the precise meaning of the term ``non-local'' is far from being unique and 
their exists a vast literature on that topic (see e.g. \cite{cushing}). A thorough 
discussion of that issue is far beyond the scope of the present paper. However, one 
can reasonably state, that the ``non-locality'' of the \dbb\ is more explicit 
(i.e. dynamical) than the ``non-separability'' of ordinary quantum mechanics\footnote{
According to fairly common usage, ''separability'' means that the {\em state} of an 
extended system can be written as a product of local states while ``locality'' expresses 
that no {\em interaction} propagates faster than light.}. 
Anyhow, for both, ordinary quantum mechanics and the \dbb, it is ensured that the ``non-locality'' or ``non-separability''
can not be used for superluminal signalling. But whether this is enough for full compatibility
between quantum mechanics and special relativity has been challenged e.g. by Ballentine \cite{ballentine}:
\begin{quote} 
However it is not clear that the requirements of special 
relativity are exhausted by excluding superluminal signals. Nor is 
it clear how one can have superluminal influences (so as to 
violate Bell's inequality and satisfy quantum mechanics) that in 
principle can not be used as signals. ($\cdots$) Whether or not there 
is a deeper incompatibility between quantum mechanics and 
relativity is not certain. 
\end{quote} 

Another indication for  ``non-locality'' in quantum mechanics is given 
if one  adopts the collapse of the  wavefunction to be a real physical process\footnote{In fact, in the context of the ``measurement
 problem'' the collapse of the wavefunction gives rise to other problems as well.}.
After all the collapse is supposed to reduce the wavefunction instantaneously and requires thereby a preferred 
frame-of-reference\footnote{An obvious solution to this problem is to suppose that the collapse occurs only along the backward 
light cone of the measurement interaction \cite{bloch,hellwig}. See e.g. \cite{barrett04} for a discussion of the problems one 
faces in this approach.} \cite{cushing_qm,maudlin}. Maudlin argues that the collapse postulate 
in combination with entangled states leads necessarily to a preferred foliation of 
space-time \cite[p.297]{maudlin}. 
While the Dirac equation provides a Lorentz covariant generalization of the Schr\"odinger equation
the satisfactory generalization of the measurement theory into the relativistic domain is still wanting.
Ironically this specific source of non-locality does not arise in the \dbb\ since here the 
collapse of the wavefunction becomes dispensable. However, as mentioned above, non-locality  figures prominently in the \dbb\ 
which makes the reconciliation with relativity challenging as well.

\subsubsection{The Bohm-Dirac theory}
Non-locality clearly provides a challenge for a satisfactory relativistic generalization of quantum mechanics or the \dbb.
However, relativistic generalizations of the \dbb\ do exist. E.g. for a Dirac particles Bohm \cite{undivided} has 
proposed the following guiding equation (the corresponding framework may be called ``Bohm-Dirac theory''):
\begin{eqnarray} 
\label{dirac}
{\mathbf v} = { \psi^{\dagger} {\bf {\mathbf \alpha}} \psi\over |\psi|^2}
\end{eqnarray}
Here $\psi$ is a solution of the Dirac equation, $\psi^{\dagger}$ its conjugate and ${\mathbf \alpha}$ a 3-vector with components 
that are build from the Pauli matrices:
\begin{eqnarray} 
\alpha_i=\left ( \begin{array}{cc}         0  & \sigma_i      \\
                                     \sigma_i & 0          \end{array} \right )  
\end{eqnarray}
The generalization to the many-particle case is straightforward \cite{undivided}[p.274].
Thus, the generalization itself is not problematic. But it is an essential property of the many-particle generalization that it  
requires a preferred reference-frame i.e. the many-particle analogue of Equ.~\ref{dirac} considers all particles at the {\em same} time. 
The predictions do nonetheless  agree with the standard theory and most important the preferred reference-frame can be made 
unobservable.

In fact, as shown in \cite{detal}, it is even possible to restore Lorentz invariance for the Bohm-Dirac theory by introducing 
additional structure. D\"urr et al. introduce an arbitrary space-like preferred slicing of space-time, determined by a Lorentz 
invariant law. An other strategy is pursued by Berndl et al. \cite{rela} who suggest a preferred joint parameterization 
(i.e. synchronization). This works provide an important step towards a Lorentz invariant \dbb\ and a counter 
example to the common 
claim that non-locality and Lorentz invariance are in strict opposition. However, these authors admit that they have not reached yet what 
Bell called ``serious Lorentz invariance'' \cite{speakable}[p.179f]. Furthermore the corresponding models  consider only entangled but 
{\em noninteracting} Dirac particles.
The relativistic generalization of the \dbb\ is also addressed in \cite{chris,GoTu03,squires}. A thorough discussion of the relation 
between non-locality and relativity can be found in \cite{dickson}.

Summing up, we have seen that non-locality and the relativistic generalization provide a challenge not only for the \dbb\ but 
also for ordinary quantum mechanics\footnote{Since the ultimate cause for non-locality is that the wavefunction of a $N$-particle 
system is defined on the configuration space, ${\rm I\!R}^{3N}$, it is not surprising that this ``non-locality'' is not a 
particular problem of the \dbb\ but for quantum mechanics in general.}. The violation of the Bell-inequality implies that the 
relation between quantum mechanics and special relativity is more subtle than customarily assumed. The concept of wavefunction 
collapse points at similar problems. 
 However, the \dbb\ does allow for a relativistic generalization when either the requirement of Lorentz invariance is relaxed 
to apply only to the observations or by introducing additional structure into the theory.

\subsection{The \dbb\ and quantum field theory}

Finally (and related to the last paragraph) there is the widespread suspicion that the concepts of the \dbb\ can not be sustained in
the realm of quantum field theories (see e.g. the letter to the editor in \cite{lte}[p. 1227] together with the reply).
However, several works on that issue have shown that there seems to be no principle problem to 
incorporate the concepts and reproduce the predictions of quantum field theories. In what follows we only sketch some of the corresponding
results. 

Similar to the situation of relativistic generalizations there are several different approaches to this question. The work on that issue 
can roughly be divided into two camps. The first (e.g. Bohm, Hiley and Holland \cite{undivided,holland}) introduces the notion of  
(bosonic-)field variables as being fundamental together with the particle position for fermions. These models provide  
laws for the evolution of these fields. However, boson like e.g. the photon do not possess a trajectory. 

The other camp (e.g. Bell \cite{speakable}[p.173] and D\"urr et al. \cite{qft1,qft2}) sustains the particle-ontology also within quantum 
field theoretical extensions of the \dbb. To this end D\"urr et al. associate the interaction part of the Hamiltonian with
jump-processes like the creation of particle-antiparticle pairs. 

While important questions remain open (see for example the discussion at the end of \cite{qft1}) it seems 
premature to reject the \dbb\ on this basis.
\section{Summary}
We have collected common criticism against the \dbb. Most of them have the merit to 
illustrate the peculiar features of this theory but they do not provide a rigorous disproof.

One strategy has been to formulate additional requirements\footnote{The requirements are 
``additional'' to the basic demand that the theory is in accordance with the 
experimental results.} which are not met by the \dbb. It remains subjective
whether this is viewed as a profound shortcoming or the radical novelty of this 
theory. After all, quantum mechanics has likewise introduced many bizarre notions into physics. 
However, while it is subjective how desirable these additional requirements are, they are clearly not irrational. 

A different strategy is to address the consistency of the \dbb\ and its ability to be generalized. The most substantial concern is 
the question of its relativistic and quantum field theoretical generalization. However, several models for such generalizations do exist
in which either the preferred foliation of space-time is unobservable or even Lorentz-invariance can be (at least formally)
sustained. Although important questions remain open it seems premature to reject the \dbb\ on this account.
Above all, these objections should be compared to those which have been advanced against other interpretations of \qm, in particular
against the orthodox view.

The merit of this discussion is to reveal that even in science a theory can not only be 
judged by its empirical confirmation\footnote{This controversy serves as a prime example for  Quine's thesis of underdetermination 
of theory by data \cite{quine}.}. In the absence of any experimental test that can 
distinguish between standard quantum mechanics and the \dbb\ one may either leave this 
question undecidable or has to invoke e.g. ``meta-theoretical'' criteria like the one presented in Sec.~\ref{meta}. 
This is completely sound but should be stated explicitly. We fully agree with Hiley who states:
\begin{quote}
Unfortunately there is a great deal of unnecessary emotion generated when
``alternative interpretations'' to quantum mechanics are discussed. By now we have so
many interpretations, that it must be clear to all that there is some basic ambiguity 
as to what the formalism is telling us about the nature of quantum processes and their 
detailed relation to those occurring in the classical domain. 
\cite{hiley}
\end{quote}
This ``unnecessary emotions'' (in part on both sides) complicate a sober discussion. 

It would be highly desirable to have an open minded discussion in the spirit of appreciation for the different interpretations.
Examples for this can be found e.g. in the camp of the \dbb, like Goldstein's work about decoherent histories \cite{page} or Tumulka's
 contribution to the GRW program \cite{rodi}. Similar the ``many-worlder'' Vaidman has made illuminative contributions to implications of
the \dbb\ in \cite{vaidman}, to pick just a few examples. However,  also the ``orthodox'' view deserves a fair discussion as expressed by 
Bell in the following less known quote:
\begin{quote}
I am not like many people I meet at conferences on the foundation of quantum
mechanics (\ldots{}) who have not really studied the orthodox theory [and] devote
their lives criticizing it (\ldots{}) I think that means they have not really
appreciated the strength of the ordinary theory. I have a very healthy respect
for it. (quoted from \cite{quantumreflection})
\end{quote}

\subsection*{Acknowledgement}
I am particularly indebted to Prof. Sheldon Goldstein for his very helpful comments and suggestions. The paper benefited greatly from 
them. Thanks also to Travis Norsen, Raymond Mackintosh, Itamar Pitowsky, Ned Floyd, Alan Forrester, Gerhard Gr\"ossing, 
Hans Dieter Zeh, Hrvoje Nikolic, Matthew Donald, Francesco Cannata, Stephan Tzenov, Giorgio Kaniadakis, Marek Czachor, Josiph Rangelov 
and Abel Miranda.   


\end{document}